\documentclass[english]{scrartcl}

\usepackage{fullpage}
\usepackage{amstext,amssymb,amsmath,amsthm}
\usepackage{graphicx,subfigure}
\usepackage{multirow}
\usepackage{url}
\usepackage{hyperref}
\usepackage{color}

\graphicspath{{.}{pics/}}

\title{Composite Bayesian inference}
\author{Alexis Roche\thanks{\url{alexis.roche@centraliens.net}}}

\def\y{{\mathbf{y}}}
\newcommand{\blambda}{{\boldsymbol{\lambda}}}

\newcommand{\bell}{{\boldsymbol{\ell}}}
\newcommand{\E}{\mathbb{E}}

\newcommand{\isep}{\mathrel{{.}\,{.}}\nobreak}

\begin{document}

\maketitle

\begin{abstract}
We revisit and generalize the concept of composite likelihood as a method to make a probabilistic inference by aggregation of multiple Bayesian agents, thereby defining a class of predictive models which we call composite Bayesian. This perspective gives insight to choose the weights associated with composite likelihood, either {\em a priori} or via learning; in the latter case, they may be tuned so as to minimize prediction cross-entropy, yielding an easy-to-solve convex problem. We argue that composite Bayesian inference is a middle way between generative and discriminative models that trades off between interpretability and prediction performance, both of which are crucial to many artificial intelligence tasks.
\end{abstract}

\section{Introduction}
\label{sec:intro}

Textbook statistical inference (frequentist or Bayesian) rests upon the existence of a probabilistic data-generating model that is both empirically valid and computationally tractable. Because this double requirement may be very challenging for multidimensional data, other inference models have been developed in applied science: deliberately misspecified generative models, as in quasi-likelihood \cite{White-82,Walker-13} or na\"ive Bayes \cite{Ng-01} methods; data compression models as in minimum description length \cite{Grunwald-07}; and discriminative models\footnote{A {\em discriminative} model is a parametric family that describes the conditional distribution of the target variable given the data, while a {\em generative} model describes the joint distribution of the target and the data.}, which currently dominate the field of artificial intelligence (AI) and typically require supervised learning on large datasets -- these include many classical machine learning \cite{Ho-95,BergerA-96,Vapnik-00,Rasmussen-06} and deep learning \cite{Lecun-15,Goodfellow-16} techniques (with the exception of deep belief networks \cite{Hinton-06}).

In a closed universe of possibilities, discriminative models can map data to predictions as well as intelligent beings or even better, however they lack introspection in the sense that they cannot {\em test} their predictions. Consider as a straightforward example deciding whether an object is a sauce pan or a frying pan based on depth. A two-class discriminative model can learn a threshold that correctly classifies most pans in practice, but will confidently classify a blender as a sauce pan, while linear discriminant analysis\footnote{Despite the name, linear discriminant analysis is based on a generative model.}, for instance, could hint that a blender is a poor match to both pan categories. Intrinsic inability to ``confirm'' or ``justify'' predictions makes discriminative models difficult to interpret.

There is growing awareness that AI should be interpretable in life-impacting applications \cite{Molnar-18}, where it is (and perhaps should remain) confined to a supportive role in human-driven decision-making processes, if only because decisions need to be explained and justified to other humans, while determining the set of possible outcomes may also be part of the process. For instance, in medical diagnosis, automated disease predictions need to be corroborated by individualized findings to be taken into account. In other words, there is little clinical value in ``black boxes'', and certainly more in automated methods that can support their predictions just like human experts. One such method that has long been used by clinicians is the familiar reference range analysis, which stems from simple generative models.

\begin{figure}[!ht]
\begin{center}
\subfigure[Generative model]{\includegraphics[width=.25\textwidth]{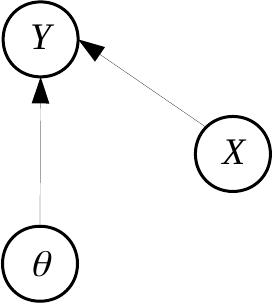}\label{fig:generative}}
\hspace*{.2\textwidth}
\subfigure[Discriminative model]{\includegraphics[width=.25\textwidth]{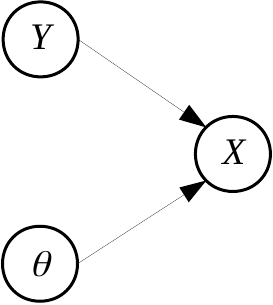}\label{fig:discriminative}}
\caption{Directed graphs representing a generative and discriminative models, where $X$, $Y$ and $\theta$ respectively denote the target variable, the data, and the model parameters. Note the marginal independence of data and parameters in the discriminative model.}
\label{fig:graph_comparison}
\end{center}
\end{figure}

A related limitation of discriminative models is that they are not suitable for unsupervised learning or on-the-fly parameter estimation because they treat the data and the model parameters as {\em marginally independent} variables, meaning that the data conveys no information about the parameters unless the target variable is observed. This is illustrated in Figure~\ref{fig:graph_comparison} by the respective directed graphs representing generative and discriminative models. For the same basic reason, supervised learning in a discriminative model is statistically less efficient than in a generative model spanning the same family of posteriors, hence it requires more training data to achieve optimal performance \cite{Ng-01}. 

This note advocates probabilistic opinion pooling \cite{Genest-86} as a natural way to merge discriminative and generative modeling approaches in order to make predictions that are both accurate and interpretable. The key idea is to combine multiple low-dimensional generative models corresponding to different pieces of information extracted from the input data. Each such model acts as an isolated ``agent'' that uses a single feature to express a testable opinion in the form of a likelihood function of the target variable. The agent opinions are then aggregated into a unique predictive probability distribution analogous to a Bayesian posterior. This idea may be understood as a probabilistic version of boosting, or as a ``divide and conquer'' approximation to the intractable Bayesian posterior. 

When choosing the aggregation method as log-linear pooling, the predictive distribution turns out to be proportional to a quantity known in computational statistics as composite likelihood (CL), see \cite{Varin-11} and the references therein. CL was developed as a proxy for traditional likelihood in parameter estimation. While maximum CL does not inherit the general property of maximum likelihood to achieve asymptotic minimum variance, it is asymptotically consistent under mild conditions \cite{Xu-11} (see Appendix~\ref{app:frequentist} for a basic weak consistency proof), and may offer an excellent trade-off between computational and statistical efficiency in practice.

Following the opinion pooling interpretation, the weights assigned to the different features in~CL may be optimized for prediction performance in a typical supervised learning scenario. As we will see, this strategy amounts to training a maximum entropy classifier using the feature-based log-likelihoods as basis functions. The likelihood functions themselves may need to be pre-trained, therefore the composite Bayesian approach typically involves a two-stage training scheme: a generative training (to learn the feature-based likelihood parameters) followed by a discriminative training (to learn the feature weights in aggregation).

The opinion pooling framework also suggests a generalization of CL, which we call {\em super composite likelihood}, in which the features may be weighted depending on target values, leading to a more flexible predictive model that can be trained similarly to CL.

The reader is warned that slightly abusive mathematical notation is sometimes used in the remainder. This is done deliberately for the sake of clarity in places where we think that notation can be lightened without creating ambiguity.

\section{Composite likelihood as opinion pooling}
\label{sec:log_pool}

Let $\mathbf{Y}$ an observable multivariate random variable with sampling distribution $p(\y|x)$ conditional on some unobserved variable of interest~$X$ taking on values in a set~${\cal X}$ assumed to be finite for simplicity. Given an experimental outcome~$\y$, the likelihood is the sampling distribution evaluated at~$\y$, seen as a function of~$x$:
$$
L(x) = p(\y|x).
$$

This requires a plausible generative model which, for complex data, may be out of reach or involve too many nuisance parameters. A natural workaround known as data reduction is to extract some lower-dimensional representation using a many-to-one mapping $z(\y)\sim f(z|x)$, and consider the potentially more convenient likelihood function:
$$
\ell(x) = f(z|x).
$$

Substituting $L(x)$ with $\ell(x)$ boils down to restricting the feature space, thereby  ``delegating'' inference to an ``agent'' provided with partial information. While it is valid for such an agent observing~$z$ only to consider $\ell(x)$ as the likelihood function of the problem, the drawback is that $\ell(x)$ might yield too vague a prediction of~$X$ due to the information loss incurred by data reduction. To make the trick statistically more efficient, we may extract several features, $z_i(\y)$ for $i=1,2,\ldots,n$, and try to combine the likelihood functions $\ell_i(x) = f(z_i|x)$ that they elicit.

If we see the likelihoods as Bayesian posterior distributions corresponding to uniform priors, this is a problem of probabilistic opinion aggregation from possibly redundant sources, for which several methods exist in the literature \cite{Tarantola-82,Genest-86,Garg-04,Allard-12}. A common choice, owing to its simplicity and tendency to produce single peaked distributions, is log-linear pooling\footnote{When $\pi(x)$ is not uniform, it is also called {\em generalized} log-linear pooling, however this terminology could be confusing here since we will present another generalization.}:
\begin{equation}
\label{eq:log_pool}
p_\blambda(x|\y) \propto \pi(x) \prod_{i=1}^n f(z_i|x)^{\lambda_i},
\end{equation} 
where $\pi(x)$ is a prior distribution and $\blambda=(\lambda_1,\ldots,\lambda_n)$ is a vector of weights. An essential property of log-linear pooling, in our opinion, is to be {\em invariant by hypothesis elimination} in the sense that  probability ratios are not altered by restricting the state space~${\cal X}$ (see Appendix~\ref{app:log_pool_standard}). Negative weights should be ruled out to warrant a no less expected monotonicity property: if all agents agree that an hypothesis $x_1$ is less probable than an hypothesis $x_2$, this should be reflected by the consensus probabilities, {\em i.e.}, we should have $p_\blambda(x_1|\y)\leq p_\blambda(x_2|\y)$. Also, we shall assume for feasibility that the weights are intrinsic to the agents and are therefore independent from the data~$\y$.

\section{Composite Bayes rule}
\label{sec:bayes_rule}

Log-linear pooling~(\ref{eq:log_pool}) bears a striking similarity to Bayes rule as it yields the form: 
$$
p_\blambda(x|\y)\propto \pi(x) L^c_\blambda(x),
$$
where the distribution $\pi(x)$ plays the role of a Bayesian prior, while the quantity:
\begin{equation}
\label{eq:comp_lik}
L^c_\blambda(x) \equiv \prod_{i=1}^n \ell_i (x)^{\lambda_i}
\end{equation} 
plays that of a likelihood function, and happens to be known in computational statistics as {\em marginal} CL \cite{Varin-11}. The slightly more general {\em conditional} CL form can be derived in the same way as above by conditioning all probabilities on confounding variables, see Appendix~\ref{app:conditional}. 

The clear computational advantage of CL over genuine likelihood is that it is more efficient to evaluate the marginal distributions of each feature than the joint distribution of all features. CL shares a convenient factorized form with the likelihood derived under the assumption of mutual feature independence, often referred to as {\em na\"ive Bayes} method in the machine learning literature, which corresponds to the special case of unitary weights, $\blambda\equiv 1$. Since log-linear pooling does not assume feature independence, it yields both a generalization and an alternative interpretation of na\"ive Bayes, whereby unitary feature weights may be used by default but do not guarantee optimal prediction performance.

\section{Composite likelihood calibration}
\label{sec:calibration}

CL involves two types of parameters:
\begin{enumerate}
    \item ``generative'' parameters, {\em i.e.}, the parameters of the feature generation models $f(z_i|x)$; 
    \item ``discriminative'' parameters, {\em i.e.},  the weights~$\blambda=(\lambda_1,\ldots,\lambda_n)$ assigned to the feature-based likelihood functions in~(\ref{eq:log_pool}).
\end{enumerate}

If training data is available, the generative parameters may be tuned in a first phase using standard estimation techniques. Otherwise, they may be considered as nuisance parameters and eliminated at prediction time by applying to CL one of the same techniques as for traditional likelihood \cite{Berger-99}. We focus in the sequel on tuning the composite weights.

\subsection{Agnostic weights}

In the absence of knowledge, uniform weights may be chosen under a rule of indifference. CL is then a scaled version of na\"ive Bayes likelihood, the unique weight value being irrelevant to the maximum CL estimator (MCLE). To get meaningful predictive probabilities, it may be tuned empirically so as to best adjust the pseudo posterior variance matrix to the asymptotic MCLE variance matrix \cite{Pauli-11}, or via a close-in-spirit curvature adjustment \cite{Ribatet-12}, as proposed in previous attempts at Bayesian inference from composite likelihood. 

Alternatively, a simple agnostic recommendation we believe to be useful for composite weights is to sum up to one. This is motivated by a characterization theorem \cite{Genest-86,Genest-86b}: the only {\em externally Bayesian} pooling operator for which the consensus probability of an outcome only depends on the agent probabilities of the same outcome, is log-linear pooling constrained to unit sum weights. The external Bayesian property is stronger than the above-mentioned invariance by hypothesis elimination; it essentially means that the consensus does not change if external opinions are taken into account before or after combining the agents. Log-linear pooling with unit sum weights also turns out to be an  optimal compromise in the sense of the inclusive Kullback-Leibler (KL) divergence \cite{Garg-04}.

Unit sum weights, however, implicitly assume maximum redundancy between features, hence tend to produce conservative predictive probabilities (see Appendix~\ref{app:frequentist}). This means low prediction confidence (over-estimated credibility sets) when features are weakly correlated, as is desirable in general. Conversely, unitary weights ($\blambda\equiv 1$) as in Na\"ive Bayes assume independent features and may thus yield over-confident predictions. An ideal method to weight features is one that can capture feature redundancy and balance it with feature relevance \cite{Peng-05}. This suggests a learning approach whenever feasible.

\subsection{Learning the weights}
\label{sec:learning}

Assuming a training dataset ${\cal D}=\{(x_k,\y_k), k=1,\ldots,N\}$ (possibly the same as for generative pre-training), the most direct way to learn the composite weights is to maximize their likelihood under the composite predictive distribution or, equivalently, minimize prediction ``cross-entropy'':
\begin{equation}
\label{eq:train_likelihood}
\max_{\blambda\succeq 0} U(\blambda),
\qquad \text{with} \quad
U(\blambda) \equiv\sum_{k=1}^N \log p_\blambda(x_k|\y_k).
\end{equation}

From~(\ref{eq:log_pool}), we see that the set of conditional distributions $p_\blambda(x|\y)$ spanned by the weights~$\blambda$ is the exponential family with natural parameter~$\blambda$ and basis functions given by the feature-based log-likelihoods:
$$
p_\blambda(x|\y) = \pi(x) \exp[\blambda^\top \bell(x,\y) - a(\blambda,y)],
$$
with:
$$
\ell_i(x,\y) \equiv \log f(z_i|x),
\qquad
a(\blambda, \y) \equiv \log \sum_{x\in{\cal X}} \pi(x) e^{\blambda^\top \bell(x,\y)}.
$$

It follows that the utility function $U(\blambda)$ in~(\ref{eq:train_likelihood}) is concave as a general property of likelihood in exponential families, hence this learning can be implemented using a standard convex optimization algorithm such as limited-memory BFGS \cite{Byrd-95}. See Appendix~\ref{app:training} for some implementation details. 

\begin{figure}[!ht]
  \begin{center}
    \includegraphics[width=.7\textwidth]{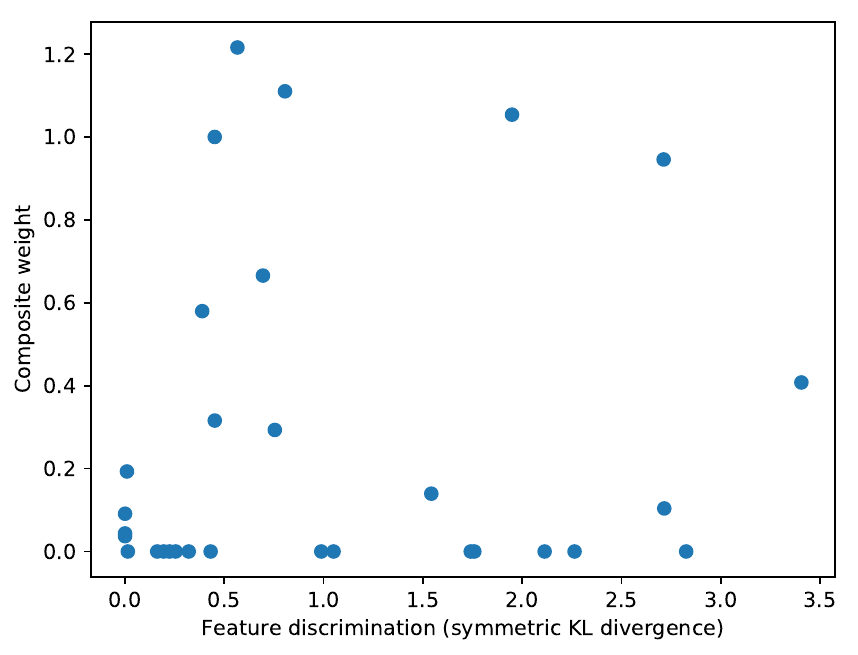}
  \end{center}
\caption{Distribution of optimal composite weights vs.~individual feature discrimination power in a binary classification task (breast cancer UCI dataset). Discrimination is measured by the maximum KL divergence between class-conditional generative distributions.}
\label{fig:disc_weight_plot}
\end{figure}

Because some positive weight constraints may turn out inactive, optimization naturally tends to produce sparse weights. As illustrated in Figure~\ref{fig:disc_weight_plot} on the breast cancer diagnostic UCI dataset \cite{Wolberg-94}, there is no clear relation between learned weights and feature-level discrimination power, showing that maximum likelihood learning differs radically from univariate feature selection as it implicitly takes feature redundancy into account via joint weight optimization.

As is customary and generally good practice, the training examples may be weighted in~(\ref{eq:train_likelihood}) so that the empirical distribution of classes matches the prior~$\pi(x)$, thus enforcing balanced classes if the prior is uniform. Also, some regularization may be needed for stability: we shall in practice maximize $U(\lambda)-\alpha \|\blambda\|^2$, where $\alpha>0$ is a small damping factor.

Other training variants that we mention here but do not particularly recommend for interpretability arise from adding basis functions to the exponential family. For instance, class-dependent offsets may be added so as to learn the ``prior'' together with the composite weights. It is also possible to perform unconstrained optimization to achieve higher training likelihood, but this means that some composite weights are then negative, in violation of the monotonicity property mentioned in Section~\ref{sec:log_pool}.

\paragraph{$I$-projection interpretation.}

Exponential family properties also imply that learning via likelihood maximization~(\ref{eq:train_likelihood}) is dual to an $I$-projection \cite{Csiszar-84}: 
\begin{equation}
\label{eq:i_proj}
\min_{p\in{\cal P}} D(p\|p_0),
\qquad \text{with} \quad
p_0(x,\y) \equiv \pi(x)h(\y),
\end{equation}
subject to the constraints $p(\y)=h(\y)$ and $\E_p[\bell] \succeq \E_h[\bell]$, where $\E_h$ denotes the expectation with respect to the joint {\em empirical} distribution of targets and inputs, $h(\y)$ is the corresponding empirical marginal distribution of inputs, and ${\cal P}$ is the set of joint probability distributions on ${\cal X}\times {\cal Y}$ with ${\cal Y}\equiv\{\y_1,\ldots,\y_N\}$. Note that the problem statement implicitly approximates the true (unknown) joint distribution of $(x,\y)$ by its empirical estimate, hence the need for regularization in practice.

The marginal constraint $p(\y)=h(\y)$ implies that only the conditional distribution $p(x|\y)$ is optimized, making the $I$-projection equivalent to maximum conditional entropy \cite{BergerA-96}. The mean log-likelihood constraints $\E_p[\bell] \succeq \E_h[\bell]$ encapsulate dependencies between the target and the data, and consider a model admissible if it guarantees the same average log-likelihood levels as observed separately for each feature. This is weaker a constraint than assuming that the feature generative models $f(z_i|x)$ are true: it only assumes that the {\em description length} \cite{Grunwald-07} achieved by each feature model is a trustworthy upper bound. Learned composite likelihood weights may be interpreted as Lagrange multipliers associated with such description length constraints. 

An insightful analysis given in \cite{Grunwald-04} expands on the fact that $I$-projection is, under broad conditions, a minimax strategy for prediction according to the log loss. Hence, any $I$-projection is in some sense a best possible predictive model given some arbitrary knowledge extracted from the training data, which is represented by mean-value constraints and is {\em inexact} in practice due to the finite training set. Trained composite likelihood only differs from other maximum entropy classifiers such as multinomial logistic regression by its underlying inexact knowledge, which in this case may be interpreted in terms of feature description length.

\section{Super composite likelihood}
\label{sec:super}

So far, we have assumed that each agent is given a fixed weight in the opinion pool~(\ref{eq:log_pool}). Let us now consider a more general pooling operator where agents can be weighted depending on the target variable:
\begin{equation}
\label{eq:super_pool}
p_\blambda(x|\y) \propto \pi(x) \prod_{i=1}^n \left[\frac{\ell_i(x)}{\ell_i(x_0)}\right]^{\lambda_i(x)},    
\end{equation}
where~$x_0\in{\cal X}$ is an arbitrary fixed reference and $\blambda: {\cal X}\to \mathbb{R}_+^n$ is now a weighting function. Note that the weights associated with the reference, $\lambda_i(x_0)$, play no role and can therefore be conventionally assumed to vanish. As shown in Appendix~\ref{app:log_pool_selective}, the main reason for introducing a reference is to maintain pooling invariance by hypothesis elimination. Moreover, (\ref{eq:super_pool}) is externally Bayesian if the weighting function sums up to one uniformly,
$$
\forall x\in{\cal X}\setminus \{x_0\},\quad
\sum_{i=1}^n \lambda_i(x) = 1,
$$
which provides a sensible default rule to tune the weights. Using the same analogy with Bayes rule as in Section~\ref{sec:bayes_rule}, we see that the quantity:
\begin{equation}
\label{eq:super_comp_lik}
L^{sc}_\blambda(x) \equiv 
\prod_{i=1}^n \left[\frac{\ell_i(x)}{\ell_i(x_0)}\right]^{\lambda_i(x)}
\end{equation}
acts as a likelihood function at prediction time. We call the form~(\ref{eq:super_comp_lik}) super composite likelihood (SCL) owing to the ``doubly composite'' aspect of combining distinctive feature-target pairs. SCL further generalizes CL since both forms are seen to be proportional in the case of binary classification ($|{\cal X}|=2$) or, generally, if the weights are constant across target values.

Conversely, if the weighting function maps each target to a unit vector ({\em i.e.}, $\forall x\in{\cal X}\setminus \{x_0\}$, $\lambda_i(x)=1$ for a single feature~$i$, and zero otherwise), then SCL turns out identical to the class-specific likelihood proxy derived in \cite{Baggenstoss-03} from a generative model selection argument -- namely, {\em PDF projection}, which, in fact, is a type of $I$-projection \cite{Minka-04}. While our derivation follows from a {\em discriminative} model selection approach, it is interesting to see that it includes the class-specific method as a special case, and thus sheds new light on this method.

A compelling advantage of SCL over CL is that it can deal with missing likelihood values, which occur if the feature generative distributions $f(z_i|x)$ are unknown for some targets~$x$. SCL makes it possible to assign zero weights $\lambda_i(x)$ to every feature-target pair for which $\ell_i(x)$ is unknown, yet using nonzero weights for other pairs. Likewise, the features for which the reference likelihood $\ell_i(x_0)$ is unknown shall receive uniformly zero weight so as to have no influence whatsoever on the predictive distribution.

If training data is available, the SCL weights may be learned by maximum likelihood as described for CL in Section~\ref{sec:learning}, yielding a similar convex problem. Since SCL enables to explore a wider class of predictive models than CL, it has the potential to achieve better prediction performance as long as overfitting does not happen during training (although SCL may not be asymptotically consistent unlike CL, see Appendix~\ref{app:frequentist}). The $I$-projection interpretation still holds, yet with class-specific mean-value constraints:
$$
\forall a\in{\cal X}\setminus\{x_0\}, \forall i\in [1\isep n],
\quad
\E_p\left[\delta_{xa}\log \frac{f(z_i|x)}{f(z_i|x_0)}\right]
\geq \E_h\left[\delta_{xa}\log \frac{f(z_i|x)}{f(z_i|x_0)}\right],
$$
where $\delta$ denotes the Kronecker delta.

\section{Discussion}
\label{sec:discussion}

The composite Bayesian framework unifies several seemingly unrelated concepts from computational statistics and machine learning: composite likelihood \cite{Varin-11}, PDF projection \cite{Baggenstoss-03}, na\"ive Bayes. It ultimately provides a rich class of trainable predictive models that work on pre-determined data features, and is therefore suitable for ``shallow'' learning tasks, where it may compete with classical discriminative models such as logistic regression, support vector machines or random forests. Potential use cases include transfer learning \cite{Goodfellow-16}, {\em i.e.}, when features are learned automatically in a related task via deep learning, for instance.

Unlike previous attempts at Bayesian inference from incomplete statistical models that relied on generative model selection \cite{Yuan-99b,Wang-14} or asymptotic theory \cite{Pauli-11,Ribatet-12}, our construction is based on a discriminative model selection approach. Composite Bayesian models are thus explicitly taylored to prediction and cannot simulate complete data but, contrary to conventional discriminative models, they encapsulate feature generation models.

Training a composite Bayesian model involves the sequential estimation of feature generative parameters and feature relevance weights. This two-stage training is reminiscent of restricted Boltzmann machines (RBMs) \cite{Hinton-06,Fischer-14}, which are fully generative models, with the important difference that it operates on {\em disjoint} sets of parameters in the case of composite Bayesian models, while RBM training modifies the same parameters twice. Also, the generative training stage of RBMs is typically unsupervised, and it is an open research issue whether composite Bayesian models are suitable for unsupervised learning.

\begin{figure}[!ht]
\begin{center}
\subfigure[Case study: clearly categorized class-3 wine]{
\includegraphics[width=\textwidth]{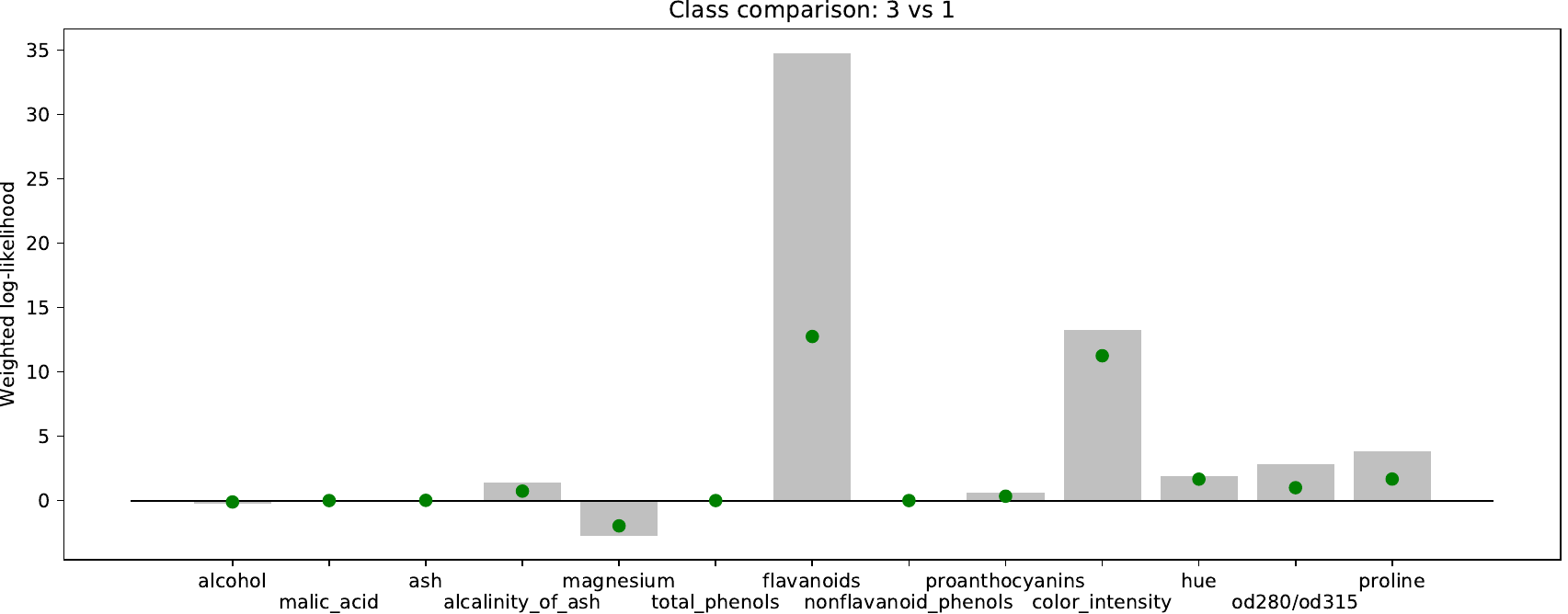}}
%\hspace*{.2\textwidth}
\subfigure[Case study: atypical class-2 wine]{
\includegraphics[width=\textwidth]{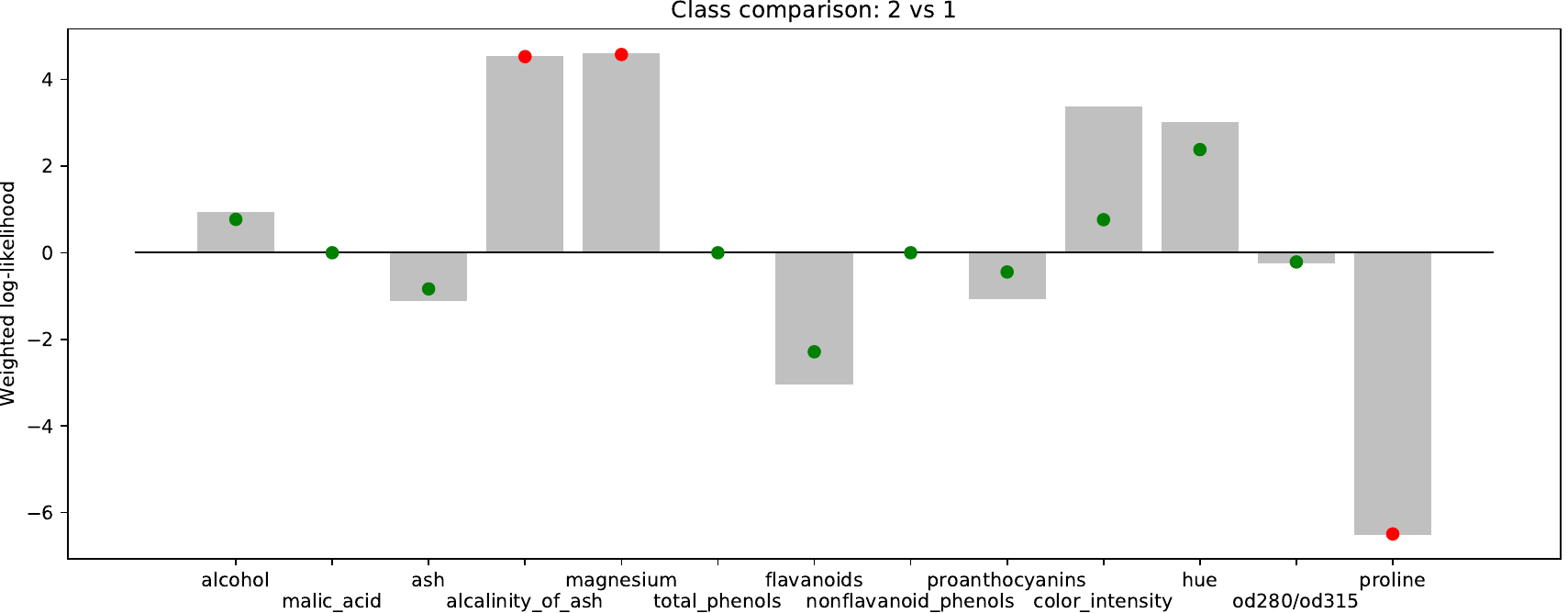}}
\end{center}
\caption{Graphical representation of feature influence on wine classification in a composite Bayesian model using heteroscedastic Gaussian feature generation models and uniform prior. Bars represent feature contributions to the difference of log-composite likelihoods between the most probable and second most probable class. Dots represent $\chi^2$ feature percentiles under the most probable class; green and red colors mark values respectively smaller and larger than 95\%.}
\label{fig:interp}
\end{figure}

Interpretability is from our point of view the key advantage of composite Bayesian models. It results from their semi-generative nature, which enables to test at prediction time how consistent the input data is with each feature distribution, thereby providing the user with crucial clues to understand and validate the prediction. Moreover, since feature log-likelihoods are combined additively, it is easy to determine which features contribute most to a particular comparison of hypotheses about the target, and summarize the rational for a prediction by highlighting a few decisive features. 

This is illustrated in Figure~\ref{fig:interp} for a composite Bayesian expert fully trained (including weight optimization as in Section~\ref{sec:learning}) on a famous UCI dataset \cite{Aeberhard-92} to classify wines into 3 categories from 13~features. The top panel shows how the expert (correctly) classifies a particular wine in category~3 with near $100\%$ probability; it can be seen that the prediction is mainly based on the flavanoid concentration and the color intensity. Since both have typical values for category~3 wines, this is a clear-cut decision. In another case depicted in the bottom panel, the wine is (again, correctly) classified in category~2 with $98.9\%$ probability, however two of the main features driving the prediction (alcanality of ash and magnesium concentration) have peculiar values for category~2, while the proline concentration quite strongly points to category~1. Despite high prediction confidence, this case may require further investigation.

We believe that there are many applications of statistical inference in which interpretation is the real goal while prediction performance is only but an intermediate requisite. A prediction method needs to prove reliable to be of any use; but, once proven reliable, it may need to be interpreted. Composite Bayesian inference may come into play when the {\em why} question is more important than the {\em what} question.

\appendix

\section{Basic frequentist properties of composite likelihood}
\label{app:frequentist}

Let a multivariate observable variable~$\y \sim p(\y|x_\star)$ distributed according to some target value~$x_\star$. For any hypothetical target value~$x$ and any extracted feature $z_i\sim f(z_i|x)$, the following double inequality holds:
$$
0 \leq
E\left[
\log \frac{f(z_i|x_\star)}{f(z_i|x)}
\right]
\leq
E\left[
\log \frac{p(\y|x_\star)}{p(\y|x)}
\right],
$$
where the expectation is taken with respect to the true distribution $p(\y|x_\star)$ at fixed~$x_\star$. This fact follows from two basic properties of KL~divergence: positivity and partition inequality.

Using a weighted sum of such inequalities, we can bracket the expected variations of the logarithm of any composite likelihood function~(\ref{eq:comp_lik}) using positive weights~$\blambda\succeq 0$:
\begin{equation}
\label{eq:variation_bound}
0 \leq
E\left[ \log \frac{L_c(x_\star, \blambda)}{L_c(x,\blambda)} \right]
\leq 
\|\blambda\|_1 E\left[ \log \frac{L(x_\star)}{L(x)} \right]
,
\end{equation}
where $\|\blambda\|_1 =\sum_i \lambda_i$ and $L(x)\equiv \log p(\y|x)$ is the true likelihood function. This trivially implies two general asymptotic properties of composite likelihood:
\begin{itemize}
\item {\em Weak consistency.} The expected composite log-likelihood is maximized by~$x_\star$.
\item {\em Conservativeness.} If the weights sum up to one or less ($\|\blambda\|_1\leq 1$), composite likelihood ratios of the true target~$x_\star$ vs.~other target values~$x$ tend to  underestimate true likelihood ratios.
\end{itemize}

Note that the derivation assumes that the weights~$\blambda$ are independent from the target~$x$, therefore these properties do not necessarily extend to SCL.

\section{Log-linear pooling}
\label{app:log_pool}

Consider a sequence of probability distributions $p_i(x)$, $i=1,\ldots,n$, representing multiple opinions on a variable of interest~$x\in{\cal X}$. The general goal of opinion aggregation is to define an operator $F(p_1,\ldots,p_n)$ that takes opinions as inputs and outputs a single distribution $p(x)$ representing a consensus.

\subsection{Standard log-linear pooling}
\label{app:log_pool_standard}

Log-linear pooling \cite{Genest-86} is one particular such operator:
$$
p(x)\propto \pi(x) \prod_{i=1}^n p_i(x)^{\lambda_i},
$$
where $\pi(x)$ is an arbitrary distribution and $\lambda_i\in\mathbb{R}$ are arbitrary weights. 

A key property of log-linear pooling is that probability ratios are stable by hypothesis elimination: if the opinions are restricted to any subset ${\cal X}_s\subset {\cal X}$ by simple renormalization, the consensus~$p_s$ on~${\cal X}_s$ is proportional to the consensus~$p$ on~${\cal X}$, {\em i.e.}, $p_s(x)\propto p(x)$ for $x\in{\cal X}_s$. In other words, eliminating some hypotheses does not bias inference towards any of the remaining hypotheses.

Another way to look at this invariance property is to state that, for any boolean-valued function $q:{\cal X}\to \{0,1\}$, opinion pooling and opinion updating commute: 
\begin{equation}
\label{eq:external_bayesian}
F[N(q p_1), \ldots, N(q p_n)]
=
N[q F(p_1,\ldots, p_n)],
\end{equation}
where~$N$ denotes the normalization endomorphism of $\mathbb{R}_+^{\cal X}$:
$$
N(f)(x) = \frac{f(x)}{\sum_{x'\in{\cal X}} f(x')}
.
$$

Under the more general condition that~(\ref{eq:external_bayesian}) holds for any non-vanishing positive valued function $q:{\cal X}\to\mathbb{R}^\star_+$,  the opinion pool is said to be {\em externally Bayesian}. It is easily seen that log-linear pooling is externally Bayesian if the weights~$\lambda_i$ sum up to one. Conversely, \cite{Genest-86b} proved that any externally Bayesian operator for which the consensus evaluated at~$x$, $F(p_1,\ldots,p_n)(x)$, is a function of $x,p_1(x),\ldots,p_n(x)$, is a log-linear pool with unit sum weights. 

\subsection{Generalization: selective pooling}
\label{app:log_pool_selective}

Log-linear pooling may not be ideal if agent have {\em selective} opinions in the sense that they respond more to certain target outcomes than to others. For instance, some agent may produce a sharply peaked distribution when the target is in a certain state but a flat distribution otherwise, and some other agents may selectively respond to other states. In such situation, a stronger consensus might be reached by weighting agents depending on the target. 

If we extend log-linear pooling by letting $\lambda_i : {\cal X}\to\mathbb{R}$ become a function of~$x$, the resulting operator,
$$
p(x)\propto \pi(x) \prod_i p_i(x)^{\lambda_i(x)},
$$
is in general not invariant by hypothesis elimination. For a subset ${\cal X}_s\subset {\cal X}$, let $p_s$ denote the consensus from opinions restricted to ${\cal X}_s$. We have, for any $x\in{\cal X}_s$,
$$
p_s(x)
\propto
p(x)
\prod_i z_i^{-\lambda_i(x)},
\qquad \text{with} \quad
z_i \equiv \sum_{x\in{\cal X}_s} p_i(x),
$$
hence the only way to have $p_s(x)$ proportional to $p(x)$ for any~${\cal X}_s$ is to impose constant weights, because the normalizing factors~$z_i$ may differ from one agent to the other.

Consider, however, the following alternative:
\begin{equation}
\label{eq:hetero_log_pool}
p(x) \propto \pi(x) \prod_i \left[\frac{p_i(x)}{p_i(x_0)}\right]^{\lambda_i(x)},    
\end{equation}
where~$x_0\in{\cal X}$ is some arbitrary reference value. Using a reference does the trick because, for any restriction~${\cal X}_s$, the odds $p_i(x)/p_i(x_0)$ are left unchanged by renormalization since the normalizing factors are the same at both the numerator and the numerator, so they cancel out. Moreover, it is easily seen that~(\ref{eq:hetero_log_pool}) is externally Bayesian whenever the weighting function verifies:
$$
\forall x \in {\cal X}\setminus \{x_0\},
\qquad
\sum_i \lambda_i(x) = 1.
$$

Clearly, (\ref{eq:hetero_log_pool}) includes log-linear pooling as a special case since the reference has no effect for a constant weighting function $\lambda_i(x)\equiv \lambda_i$. This ``selective'' generalization may be interpreted as a two-stage opinion aggregation process: first, binary classifiers operating on half overlapping sets $\{x_0,x\}$ are formed by standard log-linear pooling; second, a multi-class classifier is derived from the binary classifiers assuming preservation of odds relative to the ``pivotal'' class~$x_0$.

%However, log-linear pooling is not appropriate in situations where some agent opinions are not expressed on the whole set~${\cal X}$ of possibilities. Assume that the opinion~$p_i$ of agent~$i$ is only defined on a subset of~${\cal X}$: it could trivially be extended to~${\cal X}$ by assigning zero probability to missing outcomes, however that would mean that the probabilities provided by the agent are either underestimated or overestimated, and therefore unreliable.

\section{Conditional composite likelihood}
\label{app:conditional}

As a straightforward  extension of marginal CL, each feature-based likelihood may be conditioned by an additional ``independent'' feature $\nu_i(\y)$ considered as a predictor of the ``dependent'' feature, $z_i(\y)$, yielding the more general form:
\begin{equation}
\label{eq:cond_feat_lik}
\ell_i(x) = f(z_i|x,\nu_i).
\end{equation}

Conditioning may be useful if it is believed that $\nu_i$ alone provides little or no information about~$x$, but is informative when considered jointly with~$z_i$, as in the case of regression covariates, for instance. Equation~(\ref{eq:comp_lik}) then amounts to conditional CL \cite{Varin-11}, a more general form of CL which also includes Besag's historical {\em pseudo-likelihood} \cite{Besag-74} developed for image segmentation.

\section{Composite likelihood training}
\label{app:training}

Using the same notation as in Section~\ref{sec:learning}, the likelihood function in~(\ref{eq:train_likelihood}) can be expanded as follows:
$$
U(\blambda) 
= \sum_{k=1}^N \left[
\log \pi(x_k) + \blambda^\top \bell(x_k, \y_k) - a(\blambda,\y_k)
\right]
$$

Maximizing $U(\blambda)$ is therefore equivalent to maximizing: 
$$
\psi(\blambda) \equiv \blambda^\top \bar{\bell} - \bar{a}(\blambda), 
$$
with:
$$
\bar{\bell} \equiv \frac{1}{N} \sum_k \bell(x_k,\y_k),
\qquad
\bar{a}(\blambda) \equiv \frac{1}{N} \sum_k a(\blambda,\y_k).
$$

Note that $\psi(\blambda)$ is nothing but the dual function associated with the $I$-projection problem~(\ref{eq:i_proj}). The derivatives of~$\psi(\blambda)$ are found to be:
\begin{eqnarray*}
\nabla\psi(\blambda)
 & = & \bar{\bell} - \frac{1}{N} \sum_k \nabla a(\blambda,\y_k), \\
\nabla\nabla^\top\psi(\blambda)
 & = & - \frac{1}{N} \sum_k \nabla \nabla^\top a(\blambda,\y_k),
\end{eqnarray*}
with:
$$
\nabla a(\blambda,\y) = \E_{\blambda}(\bell),
\qquad
\nabla \nabla^\top a(\blambda,\y) = \text{Var}_{\blambda}(\bell),
$$
where $\E_{\blambda}$ and $\text{Var}_{\blambda}$ respectively denote the expectation and variance with respect to $p_\blambda(x|\y)\propto \pi(x)\exp[\blambda^\top \bell(x,\y)]$ at fixed~$\y$. This shows that $a(\blambda,\y)$ is convex in $\blambda$ since any variance matrix is positive, which, in turn, proves the concavity of~$\psi$.

%\bibliographystyle{abbrv}
%\bibliography{cvis,stat,alexis}

\end{document}